\documentstyle[12pt]{article}

\def\fo{\hbox{{1}\kern-.25em\hbox{l}}}
\def\fnote#1#2{\begingroup\def\thefootnote{#1}\footnote{#2}\addtocounter
{footnote}{-1}\endgroup}

\renewcommand{\thefootnote}{\fnsymbol{footnote}}

\def\beq{\begin{equation}}
\def\eeq{\end{equation}}
\def\eq{\end{equation}}

\begin{document}

\begin{titlepage}

\begin{flushright}
SLAC-PUB-95-6762
\end{flushright}


\LARGE
\vspace{0.3in}
\begin{center}
 Moduli Inflation from Dynamical\\ Supersymmetry
Breaking
\vspace{0.4in}

\large
Scott Thomas\fnote{\dagger}{Work supported by the Department of Energy
under contract DE-AC03-76SF00515.}\\
\vspace{.1in}
\normalsize
{\it Stanford Linear Accelerator Center\\
Stanford University\\
Stanford, CA 94309\\}

\end{center}

\vspace{0.2in}

\vspace{0.25in}

\normalsize

Moduli fields, which parameterize perturbative flat directions of
the potential in supersymmetric theories, are natural candidates
to act as inflatons.  An inflationary potential on moduli space can
result if the scale of dynamical SUSY breaking in some sector of the
theory is determined by a moduli dependent coupling.  The magnitude
of density fluctuations generated during inflation then depends on the
scale of SUSY breaking in this sector.  This can naturally be
hierarchically smaller than the Planck scale in a dynamical model,
giving small fluctuations without any fine tuning of parameters.
It is also natural for SUSY to be restored at the minimum of the
moduli potential, and to leave the universe with zero cosmological
constant after inflation.  Acceptable reheating can also be achieved
in this scenario.

\vfill
\center{\it Submitted to Physics Letters B}
\vfill

\end{titlepage}

\section{Introduction}

\baselineskip=18pt

The existence of an inflationary phase
in the early universe,
dominated by vacuum energy, eliminates the flatness and horizon problems of
standard big bang cosmology \cite{guth,infref}.
In addition, any preexisting topological defects such as
monopoles can be diluted.
Conversion of the vacuum energy
to radiation after inflation acts as the source
of entropy for our universe.
In addition, quantum deSitter fluctuations of the inflaton field(s)
driving inflation imprint a (nearly) scale invariant
spectrum of fluctuations on the background
space time metric, which can act as seeds for
structure formation.
The inflaton must be weakly coupled in order that
these fluctuations do not spoil spatial isotropy.
In fact, to be consistent with the density
and temperature fluctuations observed in the present universe,
$\delta \rho / \rho \sim \delta T / T \sim 10^{-5}$,
the inflaton potential must be extremely flat,
with a very small
dimensionless self coupling,
$\lambda \sim 10^{-8}$. 
All models of inflation must contain such a small
coupling \cite{infref}.
In this paper I outline a scheme in which
moduli act as inflatons. 
The small self coupling arises naturally from
dynamical SUSY breaking.

The requirement of introducing a small
parameter to ensure that the inflaton potential is extremely flat
makes models of inflation seem fine tuned and unnatural.
In order to be even technically natural
the couplings of the inflaton to other fields
must also be very small in non-supersymmetric theories.
Otherwise the small self coupling would not
be stable quantum mechanically.
Technical naturalness can be achieved though in
supersymmetric models \cite{susytech}.
The nonrenormalization theorem guarantees that the
superpotential is not renormalized to all orders
in perturbation theory \cite{nrt}.
The Kahler potential is renormalized but this amounts
effectively only to wave function renormalization.
The functional form of the resulting potential,
including the inflaton self coupling, is therefore
stable even if the inflaton has couplings to other fields.
This special property of supersymmetry has been
exploited to construct models of inflation
within supergravity \cite{holman},
supersymmetric GUT theories \cite{guts},
hidden sector models of supersymmetry breaking \cite{os},
and superstring theories \cite{bg,strings}.
However most of these models seem aesthetically unnatural
since the small parameter, $\lambda$,
must be input by hand to obtain
a reasonable value for $\delta \rho / \rho$.



Here I point out that 
the existence of exact perturbative flat directions
and dynamical SUSY breaking
can lead to an acceptable inflationary
potential with small self coupling.
The potential along certain directions in field space
can vanish
perturbatively in the
supersymmetric limit.
The vanishing of the perturbative potential 
for moduli makes $\lambda \ll 1$
technically natural.
More important, a nonzero moduli
potential can result from nonperturbative dynamics
which breaks supersymmetry at a scale $\mu$.
If there is a moduli dependent coupling which determines the
SUSY breaking scale $\mu$,
the moduli can act as inflatons
with vacuum energy set by the scale
$\mu^4$.
Assuming the couplings between the moduli and SUSY breaking
sector are generated at the Planck scale, $M_p$,
the moduli self coupling arises as the ratio
$\lambda \sim (\mu/M_p)^4$.  
Since the scale $\mu$ arises dynamically
by dimensional transmutation,
it can be hierarchically smaller than the
Planck scale, and a small self coupling arises naturally.
SUSY breaking in the sector responsible for driving inflation
can naturally vanish at the minimum of the moduli potential,
with zero cosmological constant,
as discussed below.
The scale $\mu$ is then in principle unrelated to
the scale of SUSY breaking responsible for the mass splittings
within the standard model supermultiplets.


In the next section I describe how moduli can act as inflatons with
a potential induced by SUSY breaking in some sector
of the theory.
Section 3 gives a simple model of dynamical SUSY breaking
which generates
a potential on moduli space.
In this model SUSY breaking and the cosmological
constant vanish at the minimum of
the moduli potential. 
The final section addresses reheating after inflation.

\section{Inflation on Moduli Space}

Supersymmetric theories can have noncompact flat directions
in field space on which the exact classical superpotential
vanishes.
The fields parameterizing such flat directions are generally
referred to as moduli.
In field theory, moduli can arise as the result
of a discrete (or continuous) $R$ symmetry.
Under a discrete $Z_N$ $R$ symmetry the superpotential
transforms as $W \rightarrow e^{4 \pi i / N} W$.
If a modulus, ${\cal M}$, is a singlet under such a symmetry,
it can not appear alone in the tree level
superpotential to any power \cite{rsym}.
An exact continuous global symmetry can also guarantee the potential
vanishes along some directions \cite{rt}.
In superstring theory world sheet symmetries can give rise
to flat directions.
For example, in (2,2) compactifications, amplitudes involving
only moduli which describe deformations of the internal
Calabi-Yau manifold vanish at zero momentum \cite{smoduli}.
These moduli therefore do not appear alone to any power in
the superpotential.
The nonrenormalization theorem guarantees
that the superpotential is not renormalized by quantum
corrections at any order
in perturbation theory \cite{nrt}.
The classical degeneracy of the potential for moduli is therefore
preserved to all orders perturbatively.
Since the perturbative potential 
vanishes for moduli, these are the natural candidates for
inflatons in supersymmetric theories \cite{bg,rt}.

It is supersymmetry which protects the moduli from obtaining
a perturbative potential.
A nontrivial potential necessary for inflation
therefore requires supersymmetry breaking.
The simplest mechanism by which a SUSY breaking potential
can be induced on moduli space is for some parameters
which describe the magnitude of
a SUSY breaking scale, $\mu$,
to depend on the moduli.
The SUSY breaking potential 
is then moduli dependent.
As discussed in the next section,
within field theory, moduli dependent couplings can
arise from the same $R$ symmetry that protects the moduli
from obtaining a potential in the supersymmetric limit.
In string theory it is common for couplings to be
moduli dependent.
It is therefore quite natural for the scale of SUSY breaking
in some sector to have moduli dependence.

If the couplings between the SUSY breaking and moduli
sectors are suppressed by the Planck scale (as would
be the case for string moduli)
the (dimensionless) parameters in the SUSY breaking sector vary by
${\cal O}(1)$ as the moduli vary by ${\cal O}(M_p)$.
The Planck mass then sets the scale for variations of
the potential on moduli space
\begin{equation}
V({\cal M}) = \mu^4 {\cal F}({\cal M}/M_p)
\label{inflatonpot}
\end{equation}
where $\mu$ is the scale of SUSY breaking during inflation,
${\cal F}$ is some model dependent function,
and $M_p = m_p / \sqrt{8 \pi}$ is the reduced
Planck Mass.
This leads to a moduli self coupling of
$\lambda \sim (\mu / M_p)^4$,
and Hubble constant during inflation of
$H \simeq ( {\cal F} / 3)^{1/2}  (\mu^2 / M_p)$.
The resulting density fluctuations are
$\delta \rho / \rho \simeq (\sqrt{75} \pi)^{-1}
({\cal F}^{3/2} / {\cal F}^{\prime}) (\mu / M_p)^2$ \cite{deltarho},
and quadrapole temperature fluctuation
$\delta T / T \simeq \sqrt{5/48} (\delta \rho / \rho)$ \cite{deltaT}.
The correct magnitude for density and temperature
fluctuations results for
$\mu \sim 10^{16}$ GeV, giving a Hubble constant
during inflation of $H \sim 10^{14}$ GeV.
During inflation the moduli kinetic energy is insignificant
and the slow roll equation gives
$\dot{{\cal M}}\simeq - H M_p ({\cal F}^{\prime}/ {\cal F})$.
The modulus acting as inflaton therefore changes by
${\cal O}(M_p)$ during one expansion time.
In order to maintain the functional form (\ref{inflatonpot})
as ${\cal M}$ evolves during inflation,
the tree level
superpotential must therefore vanish essentially to all
orders in $M_p^{-1}$ \cite{rt}.
As discussed above, this can occur as the result of
field theory or string symmetries.

Inflaton potentials with the functional form (\ref{inflatonpot}) have
been considered previously, with the scale $\mu$ input
by hand \cite{mupot}.
However, here $\mu$ is generated by nonperturbative SUSY breaking
in some sector of the theory.
Since this scale arises dynamically as the result
of dimensional transmutation, it can be hierarchically smaller
than the Planck scale.
No small parameters are input into the theory (as shown explicitly
in the next section) and a small but finite
inflaton self coupling arises naturally.
No fine tuning is required.

In order to avoid excessive SUSY breaking in the present
universe, the dynamical SUSY breaking responsible for
driving inflation in the moduli sector should vanish
at the minimum of the moduli potential.
This can arise naturally if the scale of dynamical SUSY breaking,
$\mu$, is controlled by a single moduli dependent
parameter, $\xi({\cal M})$. 
It is possible for the range of $\xi({\cal M})$
to include a value for which SUSY is unbroken.
The moduli potential then vanishes on some subspace,
$V({\cal M}_-)=0$, and
inflation ceases on ${\cal M}_-$
(assuming the cosmological  constant vanishes after inflation).
The scale of SUSY breaking should depend on a single parameter
in this scenario
since generically multiple parameters will not simultaneously vanish
on a subspace of moduli space.

It is important in this scheme
that SUSY breaking in the sector responsible for driving inflation
does in fact vanish at the minimum
of the moduli potential.
Otherwise this SUSY breaking would remain
after inflation and be transmitted
in the present universe
to the visible sector by (at least) gravitational strength
interactions.
In such a hidden sector scenario for producing the
``observed'' visible sector SUSY breaking, $\mu$ must
be identified with the intermediate scale,
$\mu \sim \sqrt{m_{3/2} M_p} \sim 10^{10-11}$ GeV,
where $m_{3/2}$ is the gravitino mass.
While this may be a natural scale for inflation
on moduli space, it leads to a Hubble constant during inflation
of $H \sim m_{3/2}$.
Such an inflationary epoch can have interesting cosmological
consequences, but generates very small density fluctuations,
$\delta \rho / \rho \sim (m_{3/2}/M_p) \sim 10^{-16}$ \cite{rt}.

\section{A Dynamical Model for the Inflaton Potential}

As an example of the scheme outlined above I consider a simple
field theory model.
The model illustrates the important feature of a single
moduli dependent parameter
which controls dynamical SUSY breaking.
For the sector of the theory which breaks supersymmetry
during inflation I take the model of Intriligator,
Seiberg, and Shenker \cite{u}.
This model contains a single matter field, $Q$, transforming
as spin $\frac{3}{2}$ under $SU(2)$.
The lowest order $SU(2)$ invariant superpotential
is nonrenormalizable
\begin{equation}
W= \frac{\beta}{M_p} Q^4
\label{Qpot}
\end{equation}
where I assume this term is generated at the Planck scale.
The Kahler potential for the
flat direction in this sector, $X=Q^4$,
has a classical singularity at the origin,
$K=(\bar{X}X)^{1/4}$.
The singularity is believed to be smoothed out for small
$X$ by nonperturbative
quantum effects, giving $K\simeq (\bar{X}X)/ \Lambda^6$,
where
$\Lambda$ is related to the $SU(2)$ dynamical scale
$\Lambda \sim \Lambda_2 = M_p e^{- 8 \pi^2 / bg^2(M_p)}$,
and $b$ is coefficient of the one loop beta function
\cite{u}.
In the presence of the superpotential (\ref{Qpot}),
this leads to supersymmetry breaking
with $Q=0$ and
vacuum energy (in the global limit) of
$V=\beta^2 \Lambda^6 / M_p^2$.
Note that for $g(M_p) \sim {\cal O}(1)$, $\Lambda$ is hierarchically
smaller than $M_p$.
No fine tuning is required to obtain a SUSY breaking scale
much less than the Planck scale.
This model amounts to nonrenormalizable SUSY
breaking since the breaking scale,
$\mu^2 \sim \Lambda^3 / M_p^2$,
vanishes in the $M_p \rightarrow 0$
limit \cite{bkn}.
Whether the SUSY breaking sector which
generates the potential on moduli space is renormalizable
or nonrenormalizable is not important.

For the moduli sector I take a single
chiral superfield ${\cal M}$.
The modulus ${\cal M}$ can be either an elementary singlet field,
or a composite field parameterizing an exact flat direction
in some other sector.
As discussed in the previous section,
tree level terms involving any power of ${\cal M}$
can be guaranteed to vanish if the superpotential is invariant
under a discrete (or continuous) $R$ symmetry, and ${\cal M}$
does not transform.
For example, the discrete $Z_3$ $R$ symmetry
$Q \rightarrow e^{i \pi/3}Q$, ${\cal M} \rightarrow {\cal M}$,
forbids self couplings of ${\cal M}$ while
\begin{equation}
W= \frac{1}{M_p} f( {\cal M}/M_p) Q^4
\label{ttw}
\end{equation}
is allowed, where $f$ is some holomorphic
function.
Over all of moduli space $Q=0$ is stable
and
$\partial W / \partial {\cal M} =0$ \cite{note}.
The potential on moduli space arises solely from the
$F$ component in the $Q$ sector,
$\partial W  / \partial Q \neq 0$.
As long as $\beta$ is in the range of $f$, there is
(at least) one point on the moduli space ${\cal M}$ for which
the vacuum energy vanishes and supersymmetry is
unbroken in the $Q$ sector.
Including supergravity interactions,
and neglecting any coupling between $Q$ and ${\cal M}$
in the Kahler potential, the potential for
${\cal M}$ with $Q=0$
is of the form (\ref{inflatonpot})
\begin{equation}
V({\cal M}) = \frac{\Lambda^6}{M_p^2}
e^{K({\cal M},\bar{{\cal M}}) / M_p^2} |\xi ( {\cal M} / M_p)|^2
\end{equation}
where $\xi=f+\beta$ and
$K({\cal M},\bar{{\cal M}})$ is the modulus Kahler potential
\cite{kahlernote}.
If $K$ and $\xi$ are such that the slow roll conditions are
satisfied, inflation can result.
A moderate amount of tuning of $K$ and $\xi$ is required in order
to obtain a sufficient number of $e$-foldings to solve the horizon
and flatness problems.
But this is true of any model of supersymmetric
inflation \cite{stewart}.
The modulus self
coupling and Hubble constant are related in this model to $\Lambda$
by $\lambda \sim (\Lambda / M_p)^6$ and
$H \sim \Lambda^3 / M_p^2$.
A dynamical scale of $\Lambda \sim 10^{16.5}$ GeV gives
the correct magnitude for density fluctuations.
Notice that since $\Lambda \gg H$ during inflation, deSitter
fluctuations do not destroy the nonperturbative effects
in the $Q$ sector.

This model has the interesting property that
the expectation value of the superpotential vanishes.
The full supergravity potential,
\break $V=e^{K/M_p^2}\left( DW \bar{D}W - 3 |W|^2/M_p^2 \right)$,
therefore vanishes at the supersymmetric minimum, leaving the universe
with zero cosmological constant after inflation.
The importance of obtaining $\langle W \rangle =0$ at a supersymmetric
stationary point ($DW=0$) has recently been emphasized
by Banks, Berkooz, and Steinhardt in the context
of the Polonyi problem \cite{banks}.
A nonzero expectation value for the superpotential
gives a negative contribution to the cosmological
constant, $-3 e^{K/M_p^2} |W|^2/M_p^2$.
If this were the case in the present context, the
universe would enter a phase of irreversible contraction
after inflation \cite{banks}.
The model above naturally avoids this problem
since the
stationary point is $Q=0$ giving $\langle W \rangle =0$
during and after inflation.
The cosmological constant therefore automatically vanishes
after inflation.

The scheme outlined above for obtaining a potential
on moduli space could be extended to other models.
For example, many dynamical models contain a single
Yukawa coupling, $h$, which controls the vacuum
energy \cite{susymodels}.
Moduli dependence of this coupling, $h({\cal M})$,
would lead to a potential on moduli space.
However in many models of this type
the supersymmetric point corresponds to
$h({\cal M}) \rightarrow \infty$.
For field theory moduli this would usually not correspond to a finite
point on moduli space.
For string theory moduli this typically represents a singular
limit.
In addition there is often a minimum
at finite $h({\cal M})$ but with SUSY broken in the moduli
sector.
Such models therefore do not seem to give acceptable
potentials for the moduli to act as inflatons.
Dynamical models in which SUSY breaking vanishes at some
finite value of a parameter (such as in the model above)
can have a SUSY preserving minimum with zero cosmological
constant at a finite point on moduli space.


\section{Reheating After Inflation}

As the inflaton evolves toward the minimum of the potential,
the slow roll conditions must eventually break down, and
inflation will cease.
The universe then 
enters an era dominated by
the coherent oscillations of the inflaton.
The inflaton eventually decays, reheating the universe.
After inflation, ${\cal M} \ll M_p$, and the inflaton oscillates in a
potential $V({\cal M}) \simeq \mu^4 | \xi({\cal M})|^2$.
If the modulus acting as inflaton is an elementary
singlet, and $\xi^{\prime}({\cal M}_-) \neq 0$,
the oscillations are harmonic with a mass $\sim \mu^2 / M_p$.
The modulus can have Planck scale suppressed
couplings to visible sector fields.
This results in a decay rate
\begin{equation}
\Gamma \sim \frac{\mu^6}{8 \pi M_p^5}
\label{singletTR}
\end{equation}
giving a reheat temperature of
$T_R \sim \sqrt{\Gamma M_p} \sim 10^{10-11}$ GeV.

One requirement for a successful inflationary
scenario is that any particles with weak scale mass
and Planck suppressed couplings not be overproduced
after inflation.
Examples of such fields include the gravitino
\cite{gravitino} and scalar moduli, usually referred to as
Polonyi fields in this context \cite{polonyi}.
The late decay of these fields can ruin the successful
predictions of nucleosynthesis.
The large reheat temperature resulting from (\ref{singletTR})
is just barely
compatible with a conservative estimate
of the bound arising from thermal production
of gravitinos and Polonyi fields \cite{gravitino,goldstino}.
However, the production of these fields
directly in the inflaton decay chain
(which may include hidden sector fields)
must also be avoided \cite{goldstino}.
The direct decay to gravitinos suffers a helicity suppression
$(m_{3/2}/M_p)^2$ as compared to (\ref{singletTR}).
The direct decay to Polonyi fields is not suppressed though.
In addition, supergravity interactions can
lead to production of Polonyi fields
by parametric resonance with the oscillating inflaton
\cite{resonant,resnote}.
Either production mechanism would lead
to a postinflationary universe eventually dominated by
nonrelativistic Polonyi fields.

This version of the Polonyi problem
is probably generic to most models of inflation, but
might be avoided in a number
of ways.
The Polonyi fields can obtain a mass from
nonperturbative dynamics not related to $m_{3/2}$ \cite{banks}.
In this case these fields either decay before
nucleosynthesis or are too heavy to be produced
in the inflaton decay.
This is perhaps natural in the context of moduli inflation
since any singlet modulus 
which couples to the SUSY breaking
sector responsible for driving inflation gains a
mass of order
$\mu^2/M_p$. 
Alternately, the modulus acting as inflaton could have larger
than Planck suppressed couplings to the visible sector.
This occurs if the inflaton is a composite flat direction
made directly of $n$ standard model fields,
${\cal M} = \phi^n$, where $\phi$ is a canonically
normalized field.
The smallest value of $n$ is 2 for the flat directions $H_u H_d$ and
$L H_u$.
As long as
$\xi^{\prime}({\cal M}_-) \neq 0$ the potential
in which such a composite inflaton oscillates after inflation
is
$V(\phi) \sim \mu^4 (\phi/M_p)^{2n}$.
Pure supergravity couplings of the composite flat direction
to other fields
give a decay rate that goes like
$\Gamma \sim \omega^3/8 \pi^2 M_p^2$,
where
$\omega \sim (\mu^2 \langle \phi \rangle^{n-1} / M_p^n)$
is the inflaton oscillation frequency.
The decay rate through such couplings
then scales like
$$
{\Gamma \over H} \sim \left({H_{\rm inf} \over M_p} \right)^2
  \left( {H \over H_{\rm inf}} \right)^{(2n-3)/n} \ll 1
$$
where $H_{\rm inf} \sim \mu^2 / M_p$
is the Hubble constant during inflation,
and $H < H_{\rm inf}$ is the time dependent
Hubble parameter after inflation.
So a composite inflaton can not efficiently decay through
Planck suppressed couplings.
In particular, it does not decay to gravitinos or Polonyi fields.
The other way the flat direction can decay is through its
Yukawa
couplings to other standard model fields.
The states coupled by a Yukawa coupling, $g$,
gain a mass $g \langle \phi \rangle$
while the flat direction is oscillating.
The ratio of the mass of these states
to the inflaton oscillation frequency is
$$
{g \langle \phi \rangle \over \omega} \sim
  g \left({M_p \over H_{\rm inf}} \right)
   \left({H_{\rm inf} \over H} \right)^{(n-2)/2}
$$
This implies that
unless $n=2$ and $g < 10^{-4}$,
$g \langle \phi \rangle > \omega$, and
decays through the Yukawa
coupling are kinematically suppressed during
this epoch.
However, when the SUSY breaking potential arising from
hidden sector breaking becomes important,
$V(\phi) \sim m_{3/2}^2 \phi^*\phi$, the flat direction can
eventually decay
through the Yukawa coupling with a reheat temperature
$T_R \sim m_{3/2} / \sqrt{g}$ \cite{rt}.
The decay of a standard model flat direction apparently does
not contribute to the Polonyi problem,
and has a reheat temperature just above the weak scale.

\section{Conclusions}

The potential of the field responsible for driving inflation
must be exceedingly flat. 
Supersymmetric moduli, with vanishing perturbative potentials,
are therefore
natural candidates to act as inflatons.
Dynamical nonperturbative SUSY breaking lifts the moduli,
giving a small but nonvanishing potential.
Moduli inflation therefore solves the naturalness and fine tuning
problems of inflation \cite{goldstone}.
It may be possible to construct intricate models in which the
scale for the potential on moduli space is related to hidden
sector SUSY breaking, while still giving acceptable density
fluctuations.
However, in the simplest models supersymmetry
is restored at the minimum of the
moduli potential.
The inflation scale is then unrelated directly to any low energy
scale.




I would like thank T. Banks, M. Dine, A. Guth, V. Rubakov, and
S. Shenker for useful discussions about inflation,
and especially R. Leigh for pointing out
the potential problem of reheating the hidden sector.
I also would like to thank
the Rutgers high energy theory group for its hospitality
during completion of this work.
During completion of this paper I became aware of
related work \cite{moduliinf} which considers the possibility
of moduli inflation.


\begin{thebibliography}{20}
\bibitem{note} The potential for small
$Q$ along the direction $Q^4$ is flat to lowest order in the global limit.
During inflation supergravity corrections lift this
local degeneracy making $Q=0$ stable.
Terms higher order in $Q^4$ in the superpotential do not
affect the SUSY breaking state at $Q=0$.
\bibitem{kahlernote} Inclusion of Kahler potential
couplings between the moduli and SUSY breaking sectors
would not change the parametric dependences in the potential.
\bibitem{resnote} Resonant production of visible sector fields is
unlikely since the large scattering cross section would destroy
coherence.
\end{thebibliography}
\end{document}